\documentstyle[12pt]{article}

\newcommand{\p}{{\partial}}
\newcommand{\cF}{{\cal F}}
\newcommand{\cG}{{\cal G}}
\newcommand{\cL}{{\cal L}}
\newcommand{\hf}{\frac12}
\newcommand{\bea}{\begin{eqnarray}}
\newcommand{\eea}{\end{eqnarray}}
\newcommand{\bean}{\begin{eqnarray*}}
\newcommand{\eean}{\end{eqnarray*}}
\newcommand{\be}{\begin{equation}}
\newcommand{\ee}{\end{equation}}
\newcommand{\ba}{\begin{array}}
\newcommand{\ea}{\end{array}}
\newcommand{\eps}{\epsilon}
\newcommand{\La}{\Lambda}

\newcommand{\la}{\langle}
\newcommand{\ra}{\rangle}
\newcommand{\del}{\delta}
\newcommand{\Del}{\Delta}
\newcommand{\om}{\omega}
\newcommand{\tr}{{\rm Tr}}

\begin{document}

\begin{titlepage}
\thispagestyle{empty}
\vspace*{-0.568in}
\begin{flushright}
LMU-TPW  96-24\\
\end{flushright}
\begin{center}
\vskip2cm
{\LARGE{ \bf Simple Calculation of Instanton Corrections 
 in Massive N=2 SU(3) SYM}}
\vskip3cm
\renewcommand{\thefootnote}{\fnsymbol{footnote}}
Holger Ewen, Kristin F\"orger~\footnote{e-mail:
hoe@mppmu.mpg.de, Kristin.Foerger@physik.uni-muenchen.de}
\renewcommand{\thefootnote}{\arabic{footnote}}
\vskip2cm
{\sl Sektion Physik, Universit\"at M\"unchen \\
Theresienstr. 37, 80333 M\"unchen, Germany}
\end{center}
\vskip3cm


\begin{abstract}
We give an explicit derivation of the Picard-Fuchs equations for $N=2$ 
supersymmetric $SU(3)$ Yang-Mills theory with  $N_f<6$ massive 
hypermultiplets in the fundamental representation. 
We determine the instanton corrections to the prepotential  
in the weak coupling region  using
the relation between $\tr\la\phi^2\ra$ and the prepotential. This method 
 can  be generalized straightforwardly to other gauge groups. 
\end{abstract}

\end{titlepage}
\newpage
\setcounter{page}{1}
\section{Introduction}
There has been much progress in duality in $N=2$ supersymmetric 
field theory as well as in string theory,  initialized by Seiberg and Witten \cite{sw}.
The main idea to solve  the effective theory by means of duality is to introduce 
a family of auxiliary curves.
The moduli space of the curves coincides with the quantum moduli 
space of the gauge theory.

In $N=2$ supersymmetric gauge theory with gauge group $G$ the low energy Wilsonian
effective action can be described by a single holomorphic prepotential $\cF$
which in $N=1$ language is a function of rank $G$ chiral multiplets $A_i$
with scalar components $a_i$.
The metric on the moduli space is $ds^2={\rm Im} \,da_{Di}d\bar a_i$
where $a_{Di}=\frac{\p\cF}{\p a_i}$ are the magnetic duals of $a_i$. 
$a_D$ and $a$ can be calculated as period integrals of a meromorphic differential on an
auxiliary complex algebraic curve.
Once we know $a_D(a)$, the prepotential can be obtained by integration and is in the 
semiclassical region of the form \cite{hkp1}:
\bea\label{prepot}
\cF_{\rm class}&=&\frac{\tau_{\rm class}}{2}\sum_{\alpha\in\Del_+} \la\alpha,a\ra^2\nonumber\\
\cF_{\rm 1-loop}&=&\frac{i}{4\pi}\,\Big\{\sum_{ \alpha\in\Del_+}\la\alpha,a\ra^2 \ln
\frac{\la\alpha,a\ra^2}{\La^2}
-\hf\sum_{i=1}^{N_f}\sum_{{w}}(\langle w,a\rangle+m_i)^2\ln\frac{(\la{w},a\ra+m_i)^2}{\La^2}\,
\Big\}\nonumber\\
\cF_{\rm inst}&=&\sum_{n=1}^\infty \cF_n(a)\La^{ (2 N_c-N_f) n }
\eea
The sums are over the positive roots $\Del_+$ of $G$ and the weights 
of the representation of the $N_f$ hypermultiplets, respectively.
In the case of $SU(3)$ the positive roots in Dynkin basis are $(2,-1)$, $(-1,2)$ and $(-1,-1)$,
and the weights of the fundamental representation $\underline{3}$ are $(1,0)$, $(0,-1)$ and $(-1,1)$.
Generalizations of this scheme to various gauge groups without \cite{klyt,af,klt} and
with matter \cite{dsh, ds,ho,aps,bl,mn,as,kp} have been given.

Although the periods and the prepotential can be obtained by explicit integration
to compute the one and two instanton corrections \cite{hkp1, hkp2},
for the $SU(2)$ case see also \cite{ms,bs},
the approach via the Picard-Fuchs (PF) equations for the periods leads also to the instanton corrections
and can easily be pushed to arbitrary orders.
PF-operators for the case of $A_r$ with $r\le 3$ without matter have been derived in \cite{klt},
a closed form for the Lie groups $A_r$, $B_r$, $C_r$ and $D_r$ has been given in \cite{a}.
The cases of massless matter for the groups with $r\le 3$ were done in \cite{iy,is,eft,imns}. 
$SU(2)$  with massive matter was investigated in \cite{oh}.

In this paper we first construct a set of PF-operators for $SU(3)$ with $N_f<6$ massive 
hypermultiplets in the fundamental representation.  From 
their  power series solutions we find their periods and 
calculate the instanton corrections to the prepotential in the weak coupling
region $u\to\infty$ using the method outlined in \cite{eft}, which relies on
the relation between the second Casimir $u$ and the prepotential \cite{m,sty,ey}.
This method should be easily applicable to other Lie groups as well.\footnote{
The Reduce procedures which were used to derive the results in this paper may be obtained by 
e-mail request from the authors.}


\section{Picard-Fuchs Operators for Massive Matter} 

We consider $N=2$ supersymmetric YM theory with gauge group $SU(3)$ and 
$N_f<6$ massive hypermultiplets in the fundamental representation.  The appropriate 
hyperelliptic curves, which define a Riemann surface of genus
two are \cite{mn}:
\be\label{cur}
y^2=W^2(x;u,v)-F(x;m,\La)
\ee
where $W=(x^3-u x-v)$ is the $A_2$ simple singularity, with
$u={\rm Tr}\langle \phi^2\rangle$
and $v={\rm Tr}\langle \phi^3\rangle$, where $\phi=\sum a_i H_i$, with $H_i$ the generators of 
the Cartan subalgebra.
The mass dependent part of the curve is:
\be\label{f}
F(x;m,\La)=\La^{2 N_c-N_f}\Big (x^{N_f}+\sum_{i=1}^{N_f} t_i x^{N_f-i}\Big)
\ee
with $t_i=\sum_{j_1<\ldots<j_i} m_{j_1}\ldots m_{j_i}$. For $N_f<5$ the $m_i$
coincide with the hypermultiplet masses $M_i$ while in the case $N_f=5$
there is a constant shift $m_i=M_i-\frac{\La}{12}$. $\La$ is meant to be $\La_{N_f}$.\footnote{
The curves derived in \cite{ho,kp} coincide with the ones taken here for $N_f=1$
and $N_f=2$ while for $N_f\geq 3$ they differ in the expression for $W(x;u,v)$.
The superscript $K$ indicates the convention taken in \cite{kp}:
$W^K=W + (\La^{2N_c-N_f}/4)\sum_{i=0}^{N_f-N_c} t_ix^{N_f-N_c-i}$, with $t_0=1$.
These curves are related to ours by an appropriate transformation of $u$, $v$, $t_i$
and $\La$. In particular the shift in $u$ for $N_f=4$ to $u^K=u+\La^2/4$ and
for $N_f=5$ to $u^K=u+\La t_1/4$ will turn out to be significant for the
comparison of the instanton corrections.}

The meromorphic differential associated to the hyperelliptic curves is \cite{aps,as}:
\be\label{diff}
\lambda=\frac{x}{2 i \pi y}\Big(W'-\hf \frac{W F'}{F}\Big)\;dx
\ee
and the periods are integrals over $\lambda$: $a_i=\int_{\alpha_i} \lambda$ and 
$a_{Di}=\int_{\beta_i}\lambda$ where $\alpha_i$ and $\beta_i$ are a symplectic basis of homology 1-cycles 
on the curves (\ref{cur}), i.e. $\alpha_i\cap\beta_j=-\beta_j\cap\alpha_i=\del_{ij}$ 
and $\alpha_i\cap\alpha_j=\beta_i\cap\beta_j=0$. 
First derivatives of $\lambda$ with
respect to $u$ and $v$  yield abelian differentials of the first kind:
\bea\label{abeldiff1}
\p_v\lambda&=&\frac{1}{2\pi i y}\;dx\nonumber\\
\p_u\lambda&=&\frac{x}{2\pi iy}\;dx
\eea

For the integrals over the abelian differentials of the first kind (\ref{abeldiff1}) we can 
find PF-operators by considering the partial derivatives 
$\p_u$, $\p_v$, $\p^2_{uu}$, $\p^2_{uv}$ and $\p^2_{vv}$  of the differentials,
producing  expressions of the form $\frac{\phi(x)}{y^n}$ where $\phi(x)$ is
some polynomial in $x$. Explicitly we get:
\bea\label{dudv}
\p_u \Big(\frac{1}{y}\Big)&=&\frac{x W}{y^3} \hspace{2cm}\p_v\Big(\frac{1}{y}\Big)=\frac{W}{y^3}\nonumber\\
\p^2_{uu}\Big(\frac{1}{y}\Big)&=&\frac{2 x^2}{y^3}+\frac{3 x^2 F}{y^5} 
\hspace{.5cm} \p^2_{uv}\Big(\frac{1}{y}\Big)=\frac{2 x}{y^3}+\frac{3 x F}{y^5}\hspace{.5cm}
\p^2_{vv}\Big(\frac{1}{y}\Big)=\frac{2}{y^3}+\frac{3 F}{y^5}
\eea
The corresponding expressions for the partial derivatives applied to
$\frac{x}{y}$ differ by one additional power of $x$ in the numerator.

In \cite{eft} we reduced these expressions with the help of two identities to a
basis of differentials:
\be\label{y1}
\frac{dx}{y},\ldots,\frac{x^4\;dx}{y}
\ee
 This basis consists of the abelian differentials of first kind $\frac{dx}{y}$, 
$\frac{x \;dx}{y}$, of second kind $\frac{x^3 \;dx}{y}$, $\frac{x^4\; dx}{y}$ and of 
third kind $\frac{x^2\;dx}{y}$. 
The identities which allow the reduction to the basis above are 
valid up to  total derivatives \cite{klt,eft}.
 To reduce the degree of $x$ in the numerator we use:
\be\label{xred}
\frac{x^k}{y^l}=\frac{x^{k-6}}{y^l} \, \frac{\Big( (l-2) x \varphi-2(k-5)\psi\Big)}{2-6 l+2 k}
\ee
where $\varphi$ and $\psi$ are to be taken from $y^2=x^6+\psi(x)$ and 
$(y^2)'=6 x^{5}+\varphi(x)$.
The formula (\ref{xred}) only works for $k\neq 3 l - 1$.
Notice that this condition means no restriction in the application to $\frac{x^k}{y}$,
with $k>4$, always allowing to reduce such monomials to the basis (\ref{y1}).
The reduction of powers of $y$ in the denominator is done by 
\be 
\frac{\phi(x)}{y^l}=\frac{1}{\Delta y^{l-2}}\Big\{a\phi+\frac{2}{l-2}(b \phi)'\Big\}
\ee
which requires the decomposition of the discriminant of the curve $\Delta_{N_f}$ into 
$\Delta_{N_f}=a(x) y^2+b(x)(y^2)'$, resulting in huge expressions 
in the cases of massive matter. 

Since the expressions (\ref{dudv}) contain $1/y^3$ and  $1/y^5$ one might as well
take into consideration a different set of basic differentials  based on  
$x^k/y^3$ or $x^k/y^5$.
The power of $1/y$ can easily be increased by multiplication with the curve:
\be\label{yinc}
\frac{\phi(x)}{y^l}=\frac{(W^2-F) \phi}{y^{l+2}}
\ee
followed by repeated application of (\ref{xred}).
Doing this we arrive at differentials of the form:
\be\label{y3}
\frac{dx}{y^3}, \ldots,\frac{x^4\;dx}{y^3}, \frac{x^8\;dx}{y^3}
\ee
or  differentials depending on $1/y^5$
\be\label{y5}
\frac{dx}{y^5}, \ldots,\frac{x^4\;dx}{y^5}, \frac{x^{14}\;dx}{y^5}
\ee
These sets represent no longer a basis of differentials but form an overcomplete set. 
The dependence among the differentials in (\ref{y3}) or  (\ref{y5}) 
corresponds just to the case where equation (\ref{xred}) fails to work.
Actually equation (\ref{xred}) allows to find  the total differential which 
gives rise to the relation  by multiplying (\ref{xred}) on both sides 
with $(2-6 l+2 k)$ and considering the case where the left hand side vanishes.
This identifies 
\be
\frac{x^3\varphi(x)-6 x^2\psi(x)}{y^3}\hspace{1cm}{\rm and}\hspace{1cm}
\frac{x^9\varphi(x)-6 x^8\psi(x)}{y^5}
\ee
as total differentials, to which the same equation (\ref{xred}) can be applied to reduce
all powers of $x$ to the ones contained in (\ref{y3}) or (\ref{y5}).
 We will denote these total differentials by $r_{N_f}$.
The actual expressions for $r_{N_f}$ depend on $N_f$.
The most convenient choice of a basic set of differentials for deriving
PF-operators turns out to be the set (\ref{y5}) since this avoids the
use of the discriminant and its decomposition.

In general, PF-operators constitute a system of partial differential 
operators of second order for abelian differentials of the first kind. 
If we  consider first the differential $\p_v\lambda$ 
the PF equations appear as
\bea\label{genpf}
0=\cL_{(i)}\int\p_v\lambda
&=& \Big(c_1^{(i)}\p^2_{uu}+c_2^{(i)}\p^2_{uv}+c_3^{(i)}\p_{vv}^2+
c_4^{(i)}\p_u+c_5^{(i)}\p_v+c_6^{(i)}\Big)\int\p_v\lambda\\
&=&\int\Big\{\big(c_1^{(i)}\p^2_{uu}+c_2^{(i)}\p^2_{uv}+c_3^{(i)}\p_{vv}^2+
c_4^{(i)}\p_u+c_5^{(i)}\p_v+c_6^{(i)}\big)\p_v\lambda+c_x^{(i)}  r_{N_f}\Big\}\nonumber
\eea
 with $i=1,2$. The coefficients $c_j^{(i)}$ are polynomial functions in $u$ and $v$
and the normalization might be chosen as $c_3^{(1)}=0$ and $c_1^{(2)}=0$.
Applying the partial derivatives to $\p_v\lambda$ and reducing the resulting 
expressions to the differentials (\ref{y5}) by the procedure described above,
i.e. by the rules (\ref{xred}) and (\ref{yinc}), we find after inclusion of the
total derivative $r_{N_f}$ two vanishing nontrivial linear combinations
\be
c_1^{(i)}\overline{\p^2_{uu}\p_v\lambda}+c_2^{(i)}\overline{\p^2_{uv}\p_v\lambda}+
c_3^{(i)}\overline{\p_{vv}^2\p_v\lambda}+
c_4^{(i)}\overline{\p_u\p_v\lambda}+c_5^{(i)}\overline{\p_v\p_v\lambda}+c_6^{(i)}\overline{\p_v\lambda}
+c_x^{(i)}  r_{N_f}=0
\ee
where overlining denotes the reduction to the basis (\ref{y5}),  yielding the two PF-operators.
Similarly we determine PF-operators $\hat{\cL}_{(i)}$ for the differential 
$\p_u\lambda$ with $\hat{\cL}_{(i)}\int\p_u\lambda=0$.

In contrast to the massless case the PF operators for the 
periods of the  meromorphic differential (\ref{diff}) are third order differential operators. 
More precisely, the section $\Pi=(\vec{a}_D,\vec{a})^T$ does not
transform irreducibly under monodromy and consequently there is no $\tilde{\cL}$
of second order with $\p_v\tilde{\cL}\Pi=0$.

In the appendix we give as an example the complete set of PF-operators for $N_f=1$.
With increasing $N_f$ the PF-operators become much larger, but factorize in the case of equal masses through
powers of $(m^3-mu+v)$ resulting in expressions which are of similar size as the ones given.

\section{Power Series Solutions and Instanton Corrections}
In the weak coupling region of the moduli space the PF-equations $\cL_{(i)}\om(u,v)$ and $\hat{\cL}_{(i)}\om(u,v)$
are known to have two pure power series solutions and two solutions including logarithms each.
To calculate the instanton corrections we apply  the method  we derived in \cite{eft}  which
requires only the power series solutions.
The instanton corrections in the semiclassical patches for $u\to \infty$ and $v\to\infty$ coincide. 
For our method we will start from the power series solutions in the semiclassical region $u\to\infty$, which
are found by an ansatz of the form
\be
\om(u,v)=u^{-k} v^l\sum_{m,n=0}^{\infty} c_{mn}(\La,t_i)\,u^{-m} v^n 
\ee
where the rational indices $(k,l)$ have to be determined as part of the solution.
They turn out to be $(\frac52,0)$ ($(\frac32,0)$ for $N_f=5$) and $(1,0)$ for $\cL_{(i)}$ and 
$(\frac12,0)$, $(2,0)$ for $\hat{\cL}_{(i)}$. 

At weak coupling we denote the two power series solutions of  the PF-operators $\cL_{(i)}$ and $\hat{\cL}_{(i)}$ by
 $\om_1$,  $\om_2$ and $\om_3$, $\om_4$, respectively.
The derivatives of the periods $a_i$ with respect to $u$ and $v$ are linear combinations of the power 
series solutions of the PF-operators. The periods $a_{Di}$ also depend on the logarithmic solutions. 
Let us consider $a_1$ first. We have:
\bea\label{duadva}
\p_v a_1&=&\rho_1\om_1+\rho_2\om_2\nonumber\\
\p_u a_1&=&\rho_3\om_3+\rho_4\om_4
\eea
where $\rho_i$ are some constants which will be determined in the following.
Equations (\ref{duadva}) give rise to an integrability condition \cite{klt} by which we can eliminate two of 
the constants, e.g. $\rho_3$ and $\rho_4$. 
Integrating the system (\ref{duadva}) we can determine $a_1(u,v)$ up to a constant. 
The resulting expression for $a_1$ still depends on $\rho_1$ and $\rho_2$.
$a_2$ is analogous but with different values for the constants $\rho_1$ and $\rho_2$. 

To proceed we introduce the expressions $u_0$, $v_0$ and $\Delta_0=4 u_0^3-27 v_0^2$:
\bea
u_0&=&a_1^2+a_2^2-a_1 a_2\nonumber\\
v_0&=&a_1 a_2 (a_1-a_2)\\
\Delta_0&=&\prod_{\alpha\in\Delta_+}\la \alpha,a\ra^2\nonumber 
\eea
where $u_0$ reproduces $u$ and $v_0$ reproduces $v$ up to higher order
corrections in $\La$. In the limit $\Lambda\to 0$ these are just the equations for the classical Casimirs and
the classical discriminant.

Now it remains to fix the coefficients $\rho_1$ and $\rho_2$. 
Inverting $u_0$ and $v_0$ in the semiclassical region around $u_0\to \infty$ we find that to leading 
order\footnote{The 
normalization taken here differs from the one in \cite{eft} due to a factor $\hf$ in $a_i$. The relation
between the conventions of that paper marked by $E$ and the ones taken here are:
$\Delta_0=\frac{\Delta_0^{E}}{4^3}$ and $v_0=-\frac{v_0^{E}}{8}$ and $u_0=\frac{u_0^E}{4}$.}
\be
a_1 = \sqrt{u_0}+\hf\frac{v_0}{u_0}+\ldots\hspace{1cm}{\rm and}\hspace{1cm}
 a_2 = \sqrt{u_0}-\hf\frac{v_0}{u_0}+\ldots
\ee
In the quantized theory the coefficients $\rho_1$ and $\rho_2$
can thus be fixed by adjusting the coefficients of $\sqrt{u}$ and $\frac{v}{u}$ 
in $a_i$ to the values $1$ and $\pm\hf$, respectively.
Indeed we find after integrating (\ref{duadva}) that the $a_i$ start with 
$\rho_1\sqrt{u}+\rho_2\frac{v}{u}+\ldots$.
This completes  the determination of  the periods in terms of the power series solutions of the 
PF-operators.

We now use the relation between $u$ and the prepotential $\cF$ derived in \cite{m,sty,ey} to calculate the
instanton corrections for the theories with  $N_f<6$ massive hypermultiplets. 
Since the prepotential $\cF$ is a homogeneous
function of weight $2$ in $m_i$, $a_i$ and $\La$ it satisfies the Euler equation 
\be
2 \cF= \La\frac{\p\cF}{\p\La}+\sum_{i=1,2}a_i\frac{\p\cF}{\p a_i} + \sum_{i=1}^{N_f}m_i\frac{\p\cF}{\p m_i}
\ee
It can be shown that 
\be\label{ladfdla}
\p_u \Big( \La \frac{\p\cF}{\p\La} \Big) = \frac{2N_c-N_f}{2\pi i}
\ee
holds even in the massive case. Integrating this formula gives $u$ up to a
function $c(v,m,\La)$. 
On the other hand, taking the derivative of (\ref{prepot}) with respect to
$\Lambda$ leads to
\be
\La\frac{\p\cF}{\p\La}=\frac{1}{2\pi i}(2 N_c-N_f) u_0-\frac{N_c}{4\pi
i}\sum_{i=1}^{N_f} m_i^2+(2 N_c-N_f)\sum_{n=1}^{\infty} \cF_n n \La^{(2 N_c-N_f) n}
\ee
Comparing the integrated equation (\ref{ladfdla}) with the equation above we get 
\be\label{uinst}
u(a) = u_0 + 2i\pi \sum_{n=1}^{\infty} \cF_n(a) n \La^{(2 N_c-N_f) n} + \La^{2 N_c-N_f} \tilde c(v,m,\La)
\ee
where we have fixed the $\Lambda$ independent part of $c(v,m,\La)$ to 
$-\frac{N_c}{4i\pi}\sum_{i=1}^{N_f}m_i^2$
by using the fact that in the classical limit $u$ and $u_0$ coincide.
The remaining part of $c$ is strongly restricted by $R$-charge
considerations.
Since $u$ has charge $4$ and only positive powers of  $m$ with charge $2$ and $v$
with charge $6$ are supposed to appear, the only possible terms are $\La^2$ and $\La m$.
These can appear for $N_f=4$ and $N_f=5$, whereas in all other cases $\tilde{c}$ must vanish.

The remaining freedom in the expression for $u$ corresponds to a shift of $\cF_1$ in both cases. 
It is just this shift by which one instanton results in the literature
\cite{hkp1,is,eft} differ.
The results in \cite{hkp1,is} were derived by explicit integration from
curves given in \cite{ho,kp} which are different from each other and from the ones we use \cite{mn}.
We find that our procedure reproduces the results starting from the
corresponding curve if we set $\tilde{c}$ to zero in each case.
The shift in $u$ is the same as the one discussed in footnote 3.

After determining the coefficients $\cG_n$ in 
\be
u - u_0(u,v;m,\La) = \sum_{n=1}^{\infty}\cG_n(a)\La^{(2 N_c-N_f)n}
\ee
the instanton corrections to the prepotential are obtained by comparing the two series.

Each individual correction is a finite expression (as opposed to an infinite series in $u$ and $v$)
 if we express it in
powers of $u_0$, $v_0$ and $\Delta_0^{-1}$. The remaining constant from integrating (\ref{duadva}) is 
immediately fixed by demanding convergence and turns out to be different from zero only for $N_f=5$.\\

\noindent For $N_f=1$ we give the one, two and three instanton correction:
\bean
\cF_1^{N_f=1}&=&\frac{1}{2\pi i}\frac{3}{4\Delta_0}\Big(-3 v_0+2 m_1 u_0\Big)\\
\cF_2^{N_f=1}&=&\frac{1}{4\pi i}\Big\{
\frac{3645v_0^2}{32\Delta_0^3}\Big(-9 v_0 m_1 + u_0^2 + 3u_0m_1^2\Big)+
\frac{9}{32\Delta_0^2}\Big(17 m_1^2 u_0-111 m_1 v_0+ u_0^2\Big)\Big\}\nonumber\\
\cF_3^{N_f=1}&=&\frac{1}{6\pi i}\Big\{
\frac{3}{64\;\Delta_0^3}\Big(770 m_1^3 u_0-11457 m_1^2 v_0+266 m_1 u_0^2-367 u_0 v_0\Big)\\
&+& \frac{2187\;v_0^2}{128\Delta_0^4}\Big(442 m_1^3 u_0-2961 m_1^2 v_0+346 m_1 u_0^2-239 u_0 v_0\Big)\\
&+& \frac{14348907 \;v_0^4}{128\;\Delta_0^5}\Big(2 m_1^3 u_0-9 m_1^2 v_0+2 m_1 u_0^2-u_0 v_0\Big)\Big\}\\
\eean

\noindent For $N_f=2$ we get ($t_1=m_1+m_2$, $t_2=m_1m_2$):
\bean
\cF_1^{N_f=2}&=&\frac{1}{2\pi i}\frac{1}{4\Delta_0}\Big(2 u_0^2-9 v_0 t_1+6 u_0 t_2\Big)\\
\cF_2^{N_f=2}&=&\frac{1}{4\pi i}\Big\{
\frac{5}{128\Delta_0}
+\frac{9}{64\Delta_0^2}\Big(2t_1^2u_0^2 - 222t_1t_2v_0 - 46 t_1v_0u_0 + 34t_2^2u_0 + 
28t_2u_0^2 + 51 v_0^2\Big)\\
&+& \frac{3645v_0^2}{128\Delta_0^3}\Big(4t_1^2u_0^2 - 36t_1t_2v_0 - 12t_1v_0u_0 + 
12t_2^2u_0 + 8t_2u_0^2 +
9v_0^2\Big)\Big\}\\
\cF_3^{N_f=2}&=&\frac{1}{6\pi i}\Big\{
\frac{3}{256\;\Delta_0^2}\Big(3 u_0+23 t_1^2+220 t_2\Big)\\
&+& \frac{3}{256\Delta_0^3}\Big(
- 1468 t_1^3 v_0 u_0 + 1064 t_1^2 t_2 u_0^2 + 27369 t_1^2 v_0^2 - 45828 t_1 t_2^2 v_0
- 25764 t_1 t_2 v_0 u_0\\
& &{} - 1504 t_1 v_0 u_0^2 + 3080 t_2^3 u_0 + 3692 t_2^2 u_0^2 
+ 48105 t_2 v_0^2 + 1643 v_0^2 u_0
\Big)\\
&+& \frac{2187v_0^2}{256\Delta_0^4}\Big(
- 478 t_1^3 v_0 u_0 + 692 t_1^2 t_2 u_0^2 + 3123 t_1^2 v_0^2 - 5922 t_1 t_2^2 v_0 - 3624 t_1 t_2 v_0 u_0\\
& &{}- 382 t_1 v_0 u_0^2 +884 t_2^3 u_0 + 944 t_2^2 u_0^2 + 3690 t_2 v_0^2 + 257 v_0^2 u_0
\Big)\\
&+& \frac{14348907v_0^4}{256\Delta_0^5}\Big(
- 2 t_1^3 v_0 u_0 +4 t_1^2 t_2 u_0^2 + 9 t_1^2 v_0^2 - 18 t_1 t_2^2 v_0 
- 12 t_1 t_2 v_0 u_0 - 2 t_1 v_0 u_0^2 \\
& &{}+ 4 t_2^3 u_0 + 4 t_2^2 u_0^2 + 9 t_2 v_0^2 + v_0^2 u_0
\Big)\Big\} 
\eean

\noindent Instanton corrections for $N_f=3$ are ($t_1=m_1+m_2+m_3$, 
$t_2=m_1m_2+m_1m_3+m_2m_3$, $t_3=m_1m_2m_3$):
\bean
\cF_1^{N_f=3}&=&\frac{1}{2\pi i}\frac{1}{4\Delta_0}\Big(-3 u_0 v_0+2 u_0^2 t_1
-9 v_0 t_2+6 u_0 t_3\Big)\\
\cF_2^{N_f=3}&=&\frac{1}{4\pi i}\Big\{\frac{1}{128\Delta_0}\Big(5t_1^2 + 6t_2 - 3u_0)\\
&+& \frac{9}{64\Delta_0^2}\Big(51 t_1^2 v_0^2 - 46 t_1 t_2 v_0 u_0 + 28 t_1 t_3 u_0^2 
- 6 t_1 v_0 u_0^2 + 2 t_2^2 u_0^2 - 222 t_2 t_3 v_0+ 156 t_2 v_0^2\\
& &{}+ 34 t_3^2 u_0 - 118 t_3 v_0 u_0 + 10 v_0^2 u_0\Big)\\
&+& \frac{3645v_0^2}{128\Delta_0^3}\Big(9 t_1^2 v_0^2 - 12 t_1 t_2 v_0 u_0 + 8 t_1 t_3 u_0^2 
- 4 t_1 v_0 u_0^2 + 4 t_2^2 u_0^2 - 36 t_2 t_3 v_0+ 18 t_2 v_0^2\\
& &{}+ 12 t_3^2 u_0 - 12 t_3 v_0 u_0 + 3 v_0^2 u_0\Big)\Big\}\\
\cF_3^{N_f=3}&=&\frac{1}{6\pi i}\Big\{\frac{1}{512 \Del_0^2}\Big( 138 t_1t_2^2 + 1320 t_1^2 t_3 + 
1512 t_2 t_3 + 18 t_1^3 u_0  + 60 t_1 t_2 u_0 + 108 t_3 u_0 \nonumber\\
& &{} - 14 t_1 u_0^2 - 201 t_1^2 v_0 - 252 t_2 v_0 + 21 u_0 v_0\Big)\\
&+& \frac{3}{512 \Del_0^3} \Big(6160 t_3^3 u_0 + 2128 t_2^2 t_3 u_0^2 + 7384 t_1 t_3^2 u_0^2
- 91656 t_2 t_3^2 v_0- 2936 t_2^3 u_0 v_0\\
& &{}- 51528 t_1 t_2 t_3 u_0 v_0 - 49956 t_3^2 u_0 v_0 - 3008 t_1^2 t_2 u_0^2 v_0 
- 3044 t_2^2 u_0^2 v_0- 19120 t_1 t_3 u_0^2 v_0\\
& &{}+54738 t_1 t_2^2 v_0^2 + 96210 t_1^2 t_3 v_0^2 + 231624 t_2 t_3 v_0^2 + 3286 t_1^3 u_0 v_0^2 
+ 28896 t_1 t_2 u_0 v_0^2\\
& &{} + 35778 t_3 u_0 v_0^2 + 1126 t_1 u_0^2 v_0^2 - 33039 t_1^2 v_0^3 - 47943 t_2 v_0^3 
- 1067 u_0v_0^3\Big)\\
&+& \frac{2187v_0^2}{512 \Del_0^4}\Big(1768 t_3^3 u_0 + 1384 t_2^2 t_3 u_0^2 + 1888 t_1 t_3^2 u_0^2 
- 11844 t_2 t_3^2 v_0 - 956 t_2^3 u_0 v_0\\
& &{} -7248 t_1 t_2 t_3 u_0 v_0 - 5136t_3^2 u_0 v_0 - 764 t_1^2 t_2 u_0^2 v_0 - 1016 t_2^2 u_0^2 v_0
- 2536 t_1 t_3 u_0^2 v_0\\
& &{} + 6246 t_1 t_2^2 v_0^2+ 7380 t_1^2 t_3 v_0^2 + 17028 t_2 t_3 v_0^2 + 514 t_1^3 u_0 v_0^2 
+ 3840 t_1 t_2 u_0 v_0^2 \\
& &{} + 2676 t_3 u_0 v_0^2 + 418t_1 u_0^2 v_0^2 - 3285 t_1^2 v_0^3 - 3852 t_2 v_0^3 -275 u_0 v_0^3\Big)\\
&+&\frac{14348907 v_0^4}{512 \Delta_0^5}\Big(8 t_3^3 u_0 + 8 t_2^2 t_3 u_0^2+ 8 t_1 t_3^2 u_0^2 
- 36 t_2 t_3^2 v_0 - 4 t_2^3 u_0 v_0 - 24 t_1 t_2 t_3 u_0 v_0\\
& &{} - 12 t_3^2 u_0 v_0- 4 t_1^2 t_2 u_0^2 v_0- 4 t_2^2 u_0^2 v_0 - 8 t_1 t_3 u_0^2 v_0 
+ 18 t_1 t_2^2 v_0^2 + 18 t_1^2 t_3 v_0^2\\
& &{} + 36 t_2 t_3 v_0^2+ 2 t_1^3 u_0 v_0^2 + 12 t_1 t_2 u_0 v_0^2+ 6 t_3 u_0 v_0^2+ 2 t_1 u_0^2 v_0^2 
- 9 t_1^2 v_0^3 - 9 t_2 v_0^3 -u_0 v_0^3\Big)\Big\}
\eean

\noindent Starting with $N_f=4$ we calculated the instanton corrections for equal masses only.
Nevertheless for the one instanton contributions we can infer the general result because in this case the
dependence on $t_i$ is linear (for equal masses: $t_1=4m$, $t_2=6m^2$, $t_3=4m^3$, $t_4=m^4$):
\bean
\cF_1^{N_f=4}&=&\frac{1}{2\pi i}\frac{1}{4\Delta_0}\Big(-2 u_0^3+18 v_0^2-3 u_0 v_0 t_1+2 u_0^2 t_2
-9 v_0 t_3 +6 u_0 t_4  \Big)\\
\cF_2^{N_f=4}&=&\frac{1}{4\pi i}\Big\{\frac{1}{128 \Delta_0}\Big(u_0^2-12 u_0 m^2 -12 v_0 m+366 m^4\Big)\\
&+&\frac{9}{64 \Delta_0^2} \Big(u_0^2 v_0^2-264 u_0^2 v_0 m^3 +200 u_0^2m^6+388 u_0 v_0^2 m^2
-1576 u_0 v_0 m^5+34 u_0 m^8\\
& &{}-240 v_0^3 m + 4650 v_0^2 m^4 -888 v_0 m^7\Big)\\
&+&\frac{3645v_0^2}{128 \Delta_0^3}\Big( u_0^2 v_0^2-112 u_0^2 v_0 m^3+112 u_0^2 m^6 
+ 84 u_0 v_0^2 m^2-336 u_0 v_0 m^5+12 u_0 m^8\\
& &{} - 36 v_0^3 m+630 v_0^2 m^4 -144 v_0 m^7\Big) \Big\}\\
\cF_3^{N_f=4}&=&\frac{1}{6\pi i}\Big\{\frac{15 m^2}{512\Delta_0}+\frac{3}{1024 \Delta_0^2}\Big(59280 m^8 
+ 14800 m^6 u_0+ 360 m^4 u_0^2 - 87792 m^5 v_0 \nonumber\\
& &{} -2480 m^3 u_0 v_0 - 40 m u_0^2 v_0 + 4092 m^2 v_0^2 + 15 u_0 v_0^2\Big)\\
&+&\frac{3}{1024 \Delta_0^3}\Big(12320 m^{12} u_0 + 156704 m^{10} u_0^2 - 733248 m^{11} v_0 
- 3248800 m^9 u_0 v_0\\
& &{} - 2431488 m^7 u_0^2 v_0+ 25314660 m^8 v_0^2 + 10245648 m^6 u_0 v_0^2 
+ 908340 m^4 u_0^2 v_0^2\\
& &{} - 24334128 m^5 v_0^3 - 1240600 m^3 u_0 v_0^3- 10552 m u_0^2 v_0^3 + 961920 m^2 v_0^4 
+ 2651 u_0 v_0^4\Big)\\
&+& \frac{2187 v_0^2}{1024 \Delta_0^4} \Big(3536 m^{12} u_0+ 66944 m^{10} u_0^2 - 94752 m^{11} v_0
-511360 m^9 u_0 v_0\\
& &{} - 500160 m^7 u_0^2 v_0 + 2299320 m^8 v_0^2+ 1203888 m^6 u_0 v_0^2 + 164400 m^4 u_0^2 v_0^2\\
& &{}  - 1952208 m^5 v_0^3- 143680 m^3 u_0 v_0^3 - 1816 m u_0^2 v_0^3+ 79236 m^2 v_0^4 
+ 293 u_0 v_0^4\Big)\\
&+& \frac{14348907 v_0^4}{1024\Delta_0^5} \Big(16 m^{12} u_0 + 352 m^{10} u_0^2 - 288 m^{11} v_0 
- 1760 m^9 u_0 v_0 - 2112 m^7 u_0^2 v_0\\
& &{} + 5940 m^8 v_0^2 + 3696 m^6 u_0 v_0^2 + 660 m^4 u_0^2 v_0^2 - 4752 m^5 v_0^3  
- 440 m^3 u_0 v_0^3 - 8 m u_0^2 v_0^3 \\
& &{} + 198 m^2 v_0^4 + u_0 v_0^4\Big)\Big\}
\eean

\noindent Finally for $N_f=5$ the instanton corrections are:
\bean
\cF_1^{N_f=5}&=&\frac{1}{2\pi i}\frac{1}{4\Delta_0}\Big(
6u_0t_5 - 9v_0t_4 + 2u_0^2t_3 - 3u_0v_0t_2 - 2u_0^3t_1 + 18v_0^2t_1 - v_0u_0^2\Big)\\
\cF_2^{N_f=5}&=&\frac{1}{4\pi i}\Big\{-\frac{15}{512}\\
&+&\frac{1}{512\Delta_0}\Big(5000 m^6 + 520 m^4 u_0+ 660 m^2 u_0^2 - 3440 m^3 v_0 
- 620 m u_0 v_0 + 227 v_0^2\Big)\\ 
&+&\frac{9}{512\Delta_0^2}\Big(272 m^{10} u_0 + 2640 m^8 u_0^2 - 8880 m^9 v_0 - 27840 m^7 u_0 v_0 
- 11424 m^5 u_0^2 v_0\\
& & + 115920 m^6 v_0^2  + 26880 m^4 u_0 v_0^2 + 840 m^2 u_0^2 v_0^2 - 33120 m^3 v_0^3 
- 520 m u_0 v_0^3 + 111 v_0^4\Big)\\
&+& \frac{10935 v_0^2}{512\Delta_0^3}\Big(16 m^{10} u_0 + 240 m^8 u_0^2 - 240 m^9 v_0 
-960 m^7 u_0 v_0 - 672 m^5 u_0^2 v_0+ 2520 m^6 v_0^2\\
& &+ 840 m^4 u_0 v_0^2 + 60 m^2 u_0^2 v_0^2 - 720 m^3 v_0^3 - 20 m u_0  v_0^3 + 3 v_0^4\Big)\Big\}\\
\cF_3^{N_f=5}&=&\frac{1}{6\pi i}\Big\{
\frac{1}{3072\Delta_0}\Big(16944 m^5 + 630 m^3 u_0 - 3585 m^2 v_0 + 340 m u_0^2 - 150 v_0 u_0\Big)\\
&+& \frac{3}{2048\Delta_0^2}\Big(349200 m^{11} + 208800 m^9 u_0 - 1973520 m^8 v_0 + 49840 m^7 u_0^2\\
& &{}- 399320 m^6 v_0 u_0 + 1270782 m^5 v_0^2 - 62000 m^4 v_0 u_0^2 + 114830 m^3 v_0^2 u_0\\
& &{}- 104715 m^2 v_0^3 + 1590 m v_0^2 u_0^2 - 497 v_0^3 u_0\Big)\\
&+& \frac{3}{2048\Delta_0^3}\Big(24640 m^{15} u_0 - 1833120 m^{14} v_0 + 508160 m^{13} u_0^2
- 13771840 m^{12} v_0 u_0\\
& &{}+ 144205560 m^{11} v_0^2 - 20073184 m^{10} v_0 u_0^2 + 122906920 m^9 v_0^2 u_0 
- 444593340 m^8 v_0^3\\ 
& &{}+ 33250440 m^7 v_0^2 u_0^2 - 80967620 m^6 v_0^3 u_0 + 129109590 m^5 v_0^4 
- 7523380 m^4 v_0^3 u_0^2\\
& &{}+ 7198430 m^3 v_0^4 u_0 - 4319415 m^2 v_0^5 + 155750 m v_0^4 u_0^2 - 25835 v_0^5 u_0\Big)\\
&+& \frac{2187v_0^2}{2048\Delta_0^4}\Big(7072 m^{15} u_0 - 236880 m^{14} v_0 + 213920 m^{13} u_0^2
- 2133040 m^{12} v_0 u_0 \\
& &{}+ 12841920 m^{11} v_0^2 - 3923920 m^{10} v_0 u_0^2 + 13533520 m^9 v_0^2 u_0
- 33050160 m^8 v_0^3 \\
& &{}+ 4667520 m^7 v_0^2 u_0^2 - 7127120 m^6 v_0^3 u_0 + 7873866 m^5 v_0^4
- 495040 m^4 v_0^3 u_0^2 \\
& &{}+ 315770 m^3 v_0^4 u_0 - 131985 m^2 v_0^5 + 2450 m v_0^4 u_0^2 - 311 v_0^5 u_0\Big)\\
&+& \frac{14348907v_0^4}{2048\Delta_0^5}\Big(32 m^{15} u_0 - 720 m^{14} v_0 
+ 1120 m^{13} u_0^2- 7280 m^{12} v_0 u_0 + 32760 m^{11} v_0^2 \\
& &{}- 16016 m^{10} v_0 u_0^2 + 40040 m^9 v_0^2 u_0- 77220 m^8 v_0^3 
+ 17160 m^7 v_0^2 u_0^2 - 20020 m^6 v_0^3 u_0 \\
& &{}+ 18018 m^5 v_0^4- 1820 m^4 v_0^3 u_0^2 + 910 m^3 v_0^4 u_0 - 315 m^2 v_0^5 
+ 10 m v_0^4 u_0^2 - v_0^5 u_0\Big)\Big\}
\eean

\pagebreak[2]

The results for different $N_f$ are related by the decoupling limit where one takes a single 
hypermultiplet
mass to infinity keeping $m\Lambda_{N_f}^{2N_c-N_f} \equiv \Lambda_{N_f-1}^{2N_c-N_f+1}$ fixed.
This amounts for the expressions above in taking from $\cF_n^{N_f}$ only the terms of order $n$ in $t_i$ 
followed by the transformations $t_1\to 1$, $t_{i+1}\to t_i$ and 
$\Lambda^{2N_c-N_f}\to\Lambda^{2N_c-N_f+1}$.

Taking the massless limit $m_i\to 0$ (or $m\to -\frac{\Lambda}{12}$ in the
case $N_f=5$) we reproduce our previous results \cite{eft}.

We checked that for $N_f=1,2,3$ the one and two instanton corrections agree with the formulas given in 
\cite{hkp1}.
For $N_f=4,5$ we find coincidence after performing the shift mentioned before.
Our results coincide with the massless one instanton calculation done in \cite{is}.

\section{Generalization to other Gauge Groups}

This method of calculating instanton corrections of  arbitrary order
from PF-operators can be generalized
to other gauge groups like  $SO(2r+1)$, $SO(2r)$ and $Sp(2r)$, starting e.g. from the curves 
given in \cite{ds,bl,as}.

To fix the linear combination of the power series solutions 
$\om_i$ to the PF-operators we look at the classical limit $\La\to 0$ 
of the curves which in all cases mentioned above contains a factor 
$\prod_{i=1}^r (x^2-a_i^2)$, where $r$ is the rank of the gauge group $G$. 
Performing a  Miura transformation we get:
\be
\prod_{i=1}^r (x^2-a_i^2)=x^{2r}-\sum_{i=1}^r u_{2i} x^{2(r-i)}
\ee
where $u_{2i}$ are the gauge invariant Casimirs of order $2i$, only in the $SO(2r)$ case 
$u_{2r}=t^2$ where $t$ is the exceptional Casimir of order $r$. 
Solving this equation in the limit $u_2\to\infty$ we obtain the leading terms of 
the periods $a_i(u)$. Comparing these expressions for $a_i(u)$ with the $\om_i$ fixes
the linear combination completely.

To derive the formula corresponding to (\ref{uinst}) for $u_2(a)$ with the above gauge groups $G$
we use the perturbative part of the prepotential given in \cite{hkp2} and the expression for the instanton 
corrections as a series in $\Lambda$. Inserting the corresponding roots and weights of the gauge group
 $G$ into (\ref{prepot}), the one loop part of $\cF^G$ is:
\bea
\cF_{\rm 1-loop}^{G}&=&\frac{i}{4\pi}\{\sum_{k< l}^r\sum_{\eps=\pm}(a_k+\eps a_l)^2\ln\frac{(a_k+\eps a_l)^2
}{\La^2}
+\xi\sum_{k=1}^r a_k^2\ln\frac{a_k^2}{\La^2}\nonumber\\
& & -\hf\sum_{k=1}^r\sum_{j=1}^{N_f}\sum_{\eps=\pm}(\eps a_k+ m_j)^2\ln\frac{(\eps a_k+ m_j)^2}{\La^2}\Big\}
\eea
where $\xi$ takes the value $1$ for $SO(2r+1)$, $0$ for $SO(2r)$ and $4$ for $Sp(2r)$. The instanton part is 
\be
\cF_{\rm inst}^G = \sum_n \cF_n \Lambda^{(2r-2+\xi- N_f)2n}
\ee
The exponent of $\Lambda$ in this series is associated to the beta function.
After taking the derivative of $\cF^G$ with respect to $\La$  we find:
\bea
u_2&=&\frac{2\pi i}{(2r-2+\xi- N_f)}\Big(\La\frac{\p\cF^G}{\p\La}+\frac{r}{2\pi i}\sum_{j=1}^{N_f}m_j^2\Big)
\nonumber\\
&=&u_0+2\pi i\sum_{n=1}\cF_n 2  n \La^{2 n(2r-2+\xi- N_f)}
\eea
where $u_0=\sum_{k=1}^r a_k^2$.
The instanton corrections to the prepotential $\cF^G$ can then be obtained by
performing the same procedure as for the gauge group  $SU(3)$. 

\section*{Acknowledgment}
This work is partially
supported by GIF-the German-Israeli Foundation for Scientific Research, 
the DFG and by the European Commission TMR program ERBFMRX-CT96-0045,
in which H.E and K.F. are associated to HU-Berlin.
We thank Stefan Theisen for introducing us to the subject and inspiring
collaboration in previous work.

\begin{appendix}
\section*{Appendix: PF-Operators for $N_f=1$}

In this appendix we give as an example the complete set of PF-operators $\cL_{(i)}\int\p_v\lambda=0$ and 
$\hat{\cL}_{(i)}\int\p_u\lambda=0$ for the case of $N_f=1$ massive hypermultiplet:
\bea
\cL_{(1)} &=&\Big(c_1^{(1)}\p^2_{uu}+c_2^{(1)}\p^2_{uv}+c_4^{(1)}\p_u+c_5^{(1)}\p_v+c_6^{(1)}\Big)
\nonumber\\
\cL_{(2)} &=&\Big(c_3^{(2)}\p_{vv}^2+c_2^{(2)}\p^2_{uv}+c_4^{(2)}\p_u+c_5^{(2)}\p_v+c_6^{(2)}\Big)
\nonumber\\
\hat{\cL}_{(1)} &=&\Big(\hat{c}_1^{(1)}\p^2_{uu}+\hat{c}_2^{(1)}\p^2_{uv}+\hat{c}_4^{(1)}\p_u+
\hat{c}_5^{(1)}\p_v+\hat{c}_6^{(1)}\Big)\\
\hat{\cL}_{(2)} &=&\Big(\hat{c}_3^{(2)}\p_{vv}^2+\hat{c}_2^{(2)}\p^2_{uv}+\hat{c}_4^{(2)}\p_u+
\hat{c}_5^{(2)}\p_v+\hat{c}_6^{(2)}\Big)
\nonumber
\eea \\

\noindent For $\cL_{(1)}$ we get:
\bean
\lefteqn{
c_1^{(1)}= 4 \Big( 9375 \La^{10} m_1^2 + 2 \La^5 ( - 550 u^3 m_1 + 750 u^2 v + 15420 u^2 m_1^3 
- 31050 u v m_1^2 - 35802 u m_1^5}\\
&&{}+ 6750 v^2 m_1 + 37665 v m_1^4 + 17496 m_1^7) + 4 ( - 1036 u^5 m_1^2 + 2088 u^4 v m_1 
+ 2472 u^4 m_1^4\\
&&{}- 1260 u^3 v^2 - 3276 u^3 v m_1^3 - 1296 u^3 m_1^6 - 8235 u^2 v^2 m_1^2 + 1296 u^2 v m_1^5 +
 17010 u v^3 m_1\\
&&{}+ 17334 u v^2 m_1^4 - 6075 v^4 - 18711 v^3 m_1^3 - 8748 v^2 m_1^6) \Big)\\
\lefteqn{
c_2^{(1)}= 625 \La^{10} (11 u m_1 - 15 v + 54 m_1^3) + 4 \La^5 (6815 u^3 m_1^2 - 18450u^2 v m_1 
- 23346 u^2 m_1^4}\\
&&{} + 12375 u v^2+ 43290 u v m_1^3 + 9720 u m_1^6- 35775 v^2 m_1^2 - 7776 v m_1^5) + 
16 (68 u^5 m_1^3\\
&&{} - 3228 u^4 v m_1^2+ 72 u^4 m_1^5 + 6192 u^3 v^2 m_1 + 7416 u^3 v m_1^4 - 3240 u^2 v^3 
- 9531 u^2 v^2 m_1^3\\
&&{} - 3888 u^2 v m_1^6- 1053 u v^3 m_1^2 + 1458 u v^2 m_1^5 + 4860 v^4 m_1 + 972 v^3 m_1^4)\\
\lefteqn{
c_4^{(1)}= 16 \Big( 5 \La^5 ( - 65 u^2 m_1 + 225 u v + 438 u m_1^3 - 1170 v m_1^2 - 324 m_1^5) +
 2 ( - 1558 u^4 m_1^2}\\
&&{} + 3174 u^3 v m_1+ 3738 u^3 m_1^4 - 2160 u^2 v^2 - 5697 u^2 v m_1^3 - 1944 u^2 m_1^6 + 
189 u v^2 m_1^2\\
&&{} + 1296 u v m_1^5 +3240 v^3 m_1+ 648 v^2 m_1^4) \Big)\\
\lefteqn{
c_5^{(1)}= 4 \Big(  - 3125 \La^{10} m_1 + 90 \La^5 (10 u^2 m_1^2 - 95 u v m_1 - 164 u m_1^4 + 
125 v^2 + 189 v m_1^3 + 108 m_1^6)}\\
&&{}+ 4 (170 u^4 m_1^3 - 5751 u^3 v m_1^2 + 180 u^3 m_1^5 + 10908 u^2 v^2 m_1 + 
13068 u^2 v m_1^4 - 6075 u v^3\\
&&{}-16740 u v^2 m_1^3 - 6804 u v m_1^6 + 2673 v^3 m_1^2 + 972 v^2 m_1^5) \Big)\\
\lefteqn{
c_6^{(1)}= 48 \Big( 25 \La^5 (3 u m_1 + 5 v - 4 m_1^3) + 2 ( - 174 u^3 m_1^2 + 362 u^2 v m_1 
+ 422 u^2 m_1^4 - 300 u v^2}\\
&&{} - 705 u v m_1^3- 216 u m_1^6 + 297 v^2 m_1^2 + 108 v m_1^5) \Big)
\eean

\noindent The second PF-operator $\cL_{(2)}$ is:
\bean
\lefteqn{
c_2^{(2)}= 28125 \La^{10} m_1 + 108 \La^5 (1025 u^2 m_1^2 - 2250 u v m_1 - 1910 u m_1^4 + 625 v^2 
+ 2200 v m_1^3}\\
&&{} + 648 m_1^6)+ 16 (32 u^5 m_1 - 240 u^4 v - 540 u^4 m_1^3 - 9792 u^3 v m_1^2 + 648 u^3 m_1^5 
+ 20520 u^2 v^2 m_1\\
&&{}+ 22680 u^2 v m_1^4 - 8100 u v^3 - 23571 u v^2 m_1^3 - 11664 u v m_1^6 - 2430 v^3 m_1^2 
+ 1458 v^2 m_1^5)\\
\lefteqn{
c_3^{(2)}= c_1^{(1)}}\\
\lefteqn{
c_4^{(2)}= 48 \Big( 75 \La^5 m_1 ( - 15 u - 2 m_1^2) + 2 (40 u^4 - 1678 u^3 m_1^2 + 3330 u^2 v m_1 
+ 3798 u^2 m_1^4 - 1350 u v^2}\\
&&{}- 4023 u v m_1^3 - 1944 u m_1^6 - 540 v^2 m_1^2 + 324 v m_1^5) \Big)\\
\lefteqn{
c_5^{(2)}= 8 \Big( 15 \La^5 ( - 50 u^2 + 20 u m_1^2 - 825 v m_1 + 198 m_1^4) + 2 (80 u^4 m_1 
- 240 u^3 v - 1350 u^3 m_1^3}\\
&&{}- 18603 u^2 v m_1^2 + 1620 u^2 m_1^5 + 39960 u v^2 m_1 + 40824 u v m_1^4 - 14175 v^3 -
 45198 v^2 m_1^3\\
&&{}- 20412 v m_1^6) \Big)\\
\lefteqn{
c_6^{(2)}= 48 \Big(  - 375 \La^5 m_1 + 2 (40 u^3 - 642 u^2 m_1^2 + 1290 u v m_1 + 1326 u m_1^4 
- 450 v^2 - 1557 v m_1^3}\\
&&{}- 648 m_1^6) \Big)
\eean

\noindent The PF-operators $\hat{\cL}_{(i)}$ for the differential $\p_u\lambda$ are: 
\bean
\lefteqn{
\hat c_1^{(1)}= 40625 \La^{15} m_1 + 40 \La^{10} ( - 100 u^3 - 75 u^2 m_1^2 - 3075 u v m_1 
+ 6102 u m_1^4 - 1125 v^2 - 2835 v m_1^3}\\
&&{}- 5832 m_1^6) + 16 \La^5 ( - 2320 u^5 m_1 + 2000 u^4 v + 22941 u^4 m_1^3 -
 28050 u^3 v m_1^2 - 45900 u^3 m_1^5\\
&&{}+ 12555 u^2 v^2 m_1 + 24246 u^2 v m_1^4 + 23328 u^2 m_1^7 + 9450 u v^3 
- 17955 u v^2 m_1^3 + 17496 u v m_1^6\\
&&{}+ 7695 v^3 m_1^2 + 14580 v^2 m_1^5) + 64 u ( - 680 u^6 m_1^2 + 1680 u^5 v m_1 
+ 1588 u^5 m_1^4 - 840 u^4 v^2\\
&&{}- 2760 u^4 v m_1^3 - 864 u^4 m_1^6 - 4230 u^3 v^2 m_1^2 + 864 u^3 v m_1^5 + 7452 u^2 v^3 m_1 
+ 10017 u^2 v^2 m_1^4\\
&&{}- 4050 u v^4 - 4212 u v^3 m_1^3 - 5832 u v^2 m_1^6 - 243 v^4 m_1^2 - 4374 v^3 m_1^5) \\
\lefteqn{
\hat c_2^{(1)}= 2 \Big( 625 \La^{15} (5 u + 39 m_1^2) + 2 \La^{10} (14600 u^3 m_1 - 15500 u^2 v 
- 63420 u^2 m_1^3 + 30000 u v m_1^2}\\
&&{}+ 119880 u m_1^5 - 30375 v^2 m_1 - 63180v m_1^4 - 69984 m_1^7) + 8 \La^5 (7090 u^5 m_1^2 
- 19020 u^4 v m_1\\
&&{}- 18690 u^4 m_1^4 + 11250 u^3 v^2 + 37380 u^3 v m_1^3 + 9720 u^3 m_1^6 + 540 u^2 v^2 m_1^2 
- 3564 u^2 v m_1^5\\
&&{}- 2430 u v^3 m_1 - 64638 u v^2 m_1^4 + 4050v^4 + 33615 v^3 m_1^3 + 34992 v^2 m_1^6) 
+ 32 (20 u^7 m_1^3 \\
&&{}- 2080 u^6v m_1^2 + 24 u^6 m_1^5 + 4380 u^5 v^2 m_1 + 4620 u^5 v m_1^4 - 2160 u^4 v^3 
- 6471 u^4 v^2 m_1^3\\
&&{} - 2592 u^4 v m_1^6 + 1404 u^3 v^3 m_1^2 + 810u^3 v^2 m_1^5 - 1377 u^2 v^4 m_1 
- 81 u^2 v^3 m_1^4 + 8505 u v^4 m_1^3 \\
&&{}- 4374 v^5 m_1^2- 4374 v^4 m_1^5) \Big)\\
\lefteqn{
\hat c_4^{(1)}= 4 \Big( 25 \La^{10} ( - 80 u^2 + 1146 u m_1^2 - 585 v m_1) + 8 \La^5 ( - 2320 u^4 m_1 
+ 2000 u^3 v - 6036 u^3 m_1^3}\\
&&{}+ 18675 u^2 v m_1^2 + 19440 u^2 m_1^5 -12150 u v^2 m_1 - 35451 u v m_1^4 - 11664 u m_1^7 
+ 2025 v^3 \\
&&{}+ 13635 v^2m_1^3+ 5832 v m_1^6) + 16 ( - 1360 u^6 m_1^2 + 3360 u^5 v m_1 + 3318 u^5 m_1^4 - 
1680 u^4 v^2 \\
&&{}- 6141 u^4 v m_1^3 - 1728 u^4 m_1^6 + 7308 u^3 v^2m_1^2 + 1350 u^3 v m_1^5 - 
10287 u^2 v^3 m_1 - 10692 u^2 v^2 m_1^4\\
&&{} + 4050 u v^4+ 18549 u v^3 m_1^3 + 5832 u v^2 m_1^6 - 7047 v^4 m_1^2 - 
2916 v^3 m_1^5) \Big)\\
\lefteqn{
\hat c_5^{(1)}= 8 \Big( 5 \La^{10} (20 u^2 m_1 - 375 u v - 744 u m_1^3 + 390 v m_1^2) + 
4 \La^5 (2840 u^4 m_1^2 - 5910 u^3 v m_1}\\
&&{}- 8400 u^3 m_1^4 + 3750 u^2 v^2 + 8574 u^2 v m_1^3 + 5184 u^2 m_1^6 - 2610 u v^2 m_1^2 + 
1620 u v m_1^5\\
&&{}+ 135 v^3 m_1 - 891 v^2 m_1^4) + 8 (40 u^6 m_1^3 - 2630 u^5 v m_1^2 + 48 u^5 m_1^5+ 
4980 u^4 v^2 m_1\\
&&{}+ 5880 u^4 v m_1^4 - 2430 u^3 v^3 - 7056 u^3 v^2 m_1^3 - 3240 u^3 v m_1^6 + 3780 u^2 v^3 m_1^2 
- 162 u^2 v^2 m_1^5\\
&&{}- 810 uv^4 m_1 - 810 u v^3 m_1^4 + 243 v^4 m_1^3) \Big)\\
\lefteqn{
\hat c_6^{(1)}= 8 \Big(  - 125 \La^{10} u + 4 \La^5 ( - 290 u^3 m_1 + 250 u^2 v - 474 u^2 m_1^3 + 
150 u v m_1^2 + 540 u m_1^5}\\
&&{} + 45 v^2 m_1- 189 v m_1^4) + 8 ( - 170 u^5 m_1^2 + 420 u^4 v m_1 + 468 u^4 m_1^4 - 210 u^3 v^2 
- 888 u^3 v m_1^3\\
&&{}- 216u^3 m_1^6 + 864 u^2 v^2 m_1^2 + 162 u^2 v m_1^5 - 270 u v^3 m_1 - 270 u v^2 m_1^4 +
 81 v^3 m_1^3) \Big)
\eean

\noindent For $\hat{\cL}_{(2)}$ we get:
\bean
\lefteqn{
\hat c_2^{(2)}= 2 \Big( 15625 \La^{15} + 10 \La^{10} (1300 u^2 m_1 - 8250 u v + 9630 u m_1^3 - 
2925v m_1^2 - 11664 m_1^5)}\\
&&{}+ 8 \La^5 (9530 u^4 m_1^2 - 18780 u^3 v m_1 - 17046 u^3 m_1^4 + 22050 u^2 v^2 - 17505 u^2 v m_1^3\\
&&{}+ 5832 u^2 m_1^6 + 7560 u v^2 m_1^2 + 42768 u v m_1^5 - 2025 v^3 m_1 + 4860 v^2 m_1^4)\\
&&{}+ 32 u (320 u^6m_1 - 160 u^5 v - 924 u^5 m_1^3 - 6024 u^4 v m_1^2 + 648 u^4 m_1^5 + 
10188 u^3 v^2 m_1\\
&&{}+ 14256 u^3 v m_1^4 - 5400 u^2 v^3 - 8991 u^2 v^2 m_1^3 -7776 u^2 v m_1^6 + 810 u v^3 m_1^2 
- 3402 u v^2 m_1^5\\
&&{}+ 1215 v^4 m_1 -2916 v^3 m_1^4) \Big)\\
\lefteqn{
\hat c_3^{(2)}= \hat c_1^{(1)}}\\
\lefteqn{
\hat c_4^{(2)}= 32 \Big( 125 \La^{10} ( - 10 u - 3 m_1^2) + 15 \La^5 (120 u^3 m_1 + 220 u^2 v - 
572u^2 m_1^3 + 465 u v m_1^2 + 162 u m_1^5}\\
&&{}- 180 v^2 m_1 + 432 v m_1^4) + 2 u ( - 6704 u^4 m_1^2 + 10752 u^3 v m_1 + 15102 u^3 m_1^4 
- 5400 u^2 v^2\\
&&{}-11745 u^2 v m_1^3 - 7776 u^2 m_1^6 + 1512 u v^2 m_1^2 - 1782 u v m_1^5 +2025 v^3 m_1 
- 4860 v^2 m_1^4) \Big)\\
\lefteqn{
\hat c_5^{(2)}= 8 \Big( 25 \La^{10} (20 u m_1 - 375 v + 36 m_1^3) + 10 \La^5 ( - 176 u^3 m_1^2 
- 192 u^2 v m_1 + 504 u^2 m_1^4 + 3150 u v^2}\\
&&{}- 6210 u v m_1^3 + 2673 v^2 m_1^2+ 4860 v m_1^5) + 16 u (320 u^5 m_1 - 160 u^4 v 
- 924 u^4 m_1^3 - 4167 u^3 v m_1^2\\
&&{}+ 648 u^3 m_1^5 + 6102 u^2 v^2 m_1 + 9504 u^2 v m_1^4 - 3375 u v^3 - 3888 u v^2 m_1^3 
- 4860 u v m_1^6 - 243 v^3 m_1^2\\
&&{}- 3645 v^2 m_1^5) \Big)\\
\lefteqn{
\hat c_6^{(2)}= 8 \Big(  - 625 \La^{10} + 30 \La^5 (70 u v - 134 u m_1^3 + 69 v m_1^2 + 108 m_1^5) +
 16 u ( - 401 u^3 m_1^2 + 438 u^2 v m_1}\\
&&{}+ 798 u^2 m_1^4 - 225 u v^2 - 360u v m_1^3 - 324 u m_1^6 - 27 v^2 m_1^2 - 243 v m_1^5) \Big)
\eean

\end{appendix}



\begin{thebibliography}{99}
\bibitem{sw} N. Seiberg, E. Witten, Nucl. Phys. B 426 (1994) 19, Nucl. Phys. B 431 (1994)  484
\bibitem{hkp1}Eric D'Hoker, I. M. Krichever, D. H. Phong,{\sl The Effective Prepotential of $N=2$ 
Supersymmetric $SU(N_c)$ Gauge Theories}, hep-th/9609041
\bibitem{klyt} A. Klemm, W. Lerche, S. Theisen, S. Yankielowicz, Phys. Lett. B 344 (1995) 169 , 
hep-th/9411048, {\sl On the Monodromies of $N=2$ 
Supersymmetric Yang-Mills Theory}, Ahrenshoop Symposium 1994: 143-155, hep-th/9412158
\bibitem{af} P. C. Argyres, A. E. Faraggi, Phys. Rev. Lett. 74 (1995) 3931
\bibitem{klt} A. Klemm, W. Lerche, S. Theisen, Int. J. Mod. Phys. A 11 (1996) 1929, 
hep-th/9505150
\bibitem{dsh} M. Douglas, S. Shenker, Nucl. Phys.  B 447 (1995) 271, 
hep-th/9503163
\bibitem{ds} U. Danielson, B. Sundborg, Phys. Lett.  B 358 (1995) 273,  
hep-th/9504102
\bibitem{ho} A. Hanany, Y. Oz, Nucl. Phys.  B 452 (1995) 283, 
hep-th/9505075
\bibitem{aps} P. C. Argyres, M. R. Plesser,  A. D. Shapere, 
Phys. Rev. Lett. 75 (1995) 1699, hep-th/9505100 
\bibitem{bl}A. Brandhuber, K. Landsteiner, Phys. Lett.  B 358 (1995) 73 
\bibitem{mn} J. A. Minahan, D. Nemeschansky, Phys. Lett. B 464 (1996) 3, 
hep-th/9507032
\bibitem{as} P. C. Argyres, A. D. Shapere, Nucl. Phys. B 461 (1996) 437, hep-th/9509175
\bibitem{kp}I. M. Krichever, D. H. Phong, {\sl On the integrable geometry of soliton equations 
and N=2 supersymmetric gauge theories}, hep-th/9604199
\bibitem{hkp2}Eric D'Hoker, I. M. Krichever, D. H. Phong,{\sl The Effective Prepotential of $N=2$ 
Supersymmetric  $SO(N_c)$ and $Sp(N_c)$ Gauge Theories}, hep-th/9609145
\bibitem{ms} T. Masuda, H. Suzuki, {\sl Periods and Prepotential of $N=2$ $SU(2)$ Supersymmetric 
Yang-Mills Theory with Massive Hypermultiplets}, hep-th/9609066;
{\sl Prepotential of $N=2$ Supersymmetric Yang-Mills Theories in the Weak Coupling Region}, 
hep-th/9609065
\bibitem{bs} A. Brandhuber, S. Stieberger, {\sl Periods, Coupling Constants and Modular 
Functions in $N=2$ $SU(2)$ SYM with Massive Matter}, hep-th/9609130
\bibitem{a} M. Alishahiha, {\sl On the Picard-Fuchs equations of the SW models}, hep-th/9609157
\bibitem{iy} K. Ito, S. Yang, Phys. Lett. B 366 (1996) 165, 
hep-th/9507144, {\sl Picard-Fuchs Equations and Prepotentials in $N=2$ Supersymmetric QCD}, 
hep-th/9603073
\bibitem{is}K. Ito, N. Sasakura, Phys. Lett. B 382 (1996) 95, hep-th/9602073; 
{\sl Exact and Microscopic One-Instanton Calculations in $N=2$ 
Supersymmetric Yang-Mills Theories}, hep-th/9608054;
{\sl One-instanton calculations in $N=2$ $SU(N_c)$ 
Supersymmetric QCD}, hep-th/9609104
\bibitem{eft} H. Ewen, K. F\"orger, S. Theisen, {\sl Prepotential in $N=2$ Supersymmetric $SU(3)$ 
YM-Theory with
Massless Hypermultiplets}, hep-th/9609062
\bibitem{imns}J. M. Isidro, A. Mukherjee, J. P. Nunes, H. J. Schnitzer, {\sl A New Derivation of the 
Picard-Fuchs Equations for Effective $N=2$ Super Yang-Mills Theories}, hep-th/9609116
\bibitem{oh} Y. Ohta, {\sl Prepotentials of $N=2$ $SU(2)$ Yang-Mills Gauge 
Theory Coupled with a Massive Matter Multiplet}, hep- th/9604051, 
{\sl   Prepotentials of $N=2$ $SU(2)$ Yang-Mills Theories Coupled with Massive Matter Multiplets}, 
hep-th/9604059
\bibitem{m} M. Matone, Phys. Lett.  B357(1995)342; hep-th/9506102
\bibitem{sty} J. Sonnenschein, S. Theisen, S. Yankielowicz, Phys. Lett. B 
367 (1996) 145 , hep-th/9510129
\bibitem{ey} T. Eguchi, S. Yang, {\sl Prepotentials of $N=2$ Supersymmetric Gauge 
Theories and Soliton Equations}, hep-th/9510183 
\end{thebibliography}
\end{document}